\documentclass[conference]{IEEEtran}
\IEEEoverridecommandlockouts
\usepackage{cite}
\usepackage{amsmath,amssymb,amsfonts}
\usepackage{algorithmic}
\usepackage{graphicx}
\usepackage{textcomp}
\usepackage{xcolor}
\usepackage{mathrsfs}
\usepackage{mathtools}
\usepackage{stmaryrd}

\usepackage{tikz}

\usepackage{amsthm}

\newcommand{\set}[1]{\mathcal #1} 
\renewcommand{\vec}[1]{\boldsymbol{#1}} 
\newcommand{\mat}[1]{\boldsymbol{\underline{#1}}} 
\newcommand{\prob}[2]{\mathbb{P}_{#2}\left[#1\right]} 
\newcommand{\Exp}[2]{\mathbb{E}_{#2}\left[#1\right]} 
\newcommand{\Cov}[2]{\mathbb{C}_{#2}\left[#1\right]} 
\newtheorem{theorem}{Theorem}
\newtheorem{lemma}{Lemma}

\newtheorem{corollary}{Corollary}
\newtheorem{proposition}{Proposition}
\newtheorem{definition}{Definition}

\def\BibTeX{{\rm B\kern-.05em{\sc i\kern-.025em b}\kern-.08em
    T\kern-.1667em\lower.7ex\hbox{E}\kern-.125emX}}
\begin{document}

\title{Rate-Loss Regions for Polynomial Regression with Side Information}

\author{
 Jiahui Wei$^{1,2}$, Philippe Mary$^{2}$, and Elsa Dupraz$^{1}$ \\
\small $^{1}$ IMT Atlantique, CNRS UMR 6285, Lab-STICC, Brest, France \\ $^{2}$ Univ. Rennes, INSA, IETR, UMR CNRS, Rennes, France \thanks{ This work has received a French government support granted to the Cominlabs excellence laboratory and managed by the National Research Agency in the ``Investing for the Future'' program under reference ANR-10-LABX-07-01. This work was also funded by the Brittany region.}
}

\maketitle

\tikzset{
box/.style ={
rectangle, 
rounded corners =5pt, 
minimum width =50pt, 
minimum height =20pt, 
inner sep=5pt, 
draw=black },
bigbox/.style ={
rectangle, 
minimum width =50pt, 
minimum height =65pt, 
inner sep=5pt, 
draw=black 
},
roundnode/.style={
circle, draw=black, 
very thick, 
minimum size=6mm},
squarenode/.style={
rectangle, 
draw=black, 
very thick, 
minimum size=7mm}
}

\begin{abstract}
In the context of goal-oriented communications, this paper addresses the achievable rate versus generalization error region of a learning task applied on compressed data. The study focuses on the distributed setup where a source is compressed and transmitted through a noiseless channel to a receiver performing polynomial regression, aided by side information available at the decoder. The paper provides the asymptotic rate generalization error region, and extends the analysis to the non-asymptotic regime.
Additionally, it investigates the asymptotic trade-off between polynomial regression and data reconstruction under communication constraints. The proposed achievable scheme is shown to achieve the minimum generalization error as well as the optimal rate-distortion region. 
\end{abstract}

\begin{IEEEkeywords}
Information theory, source coding, statistical learning, rate-distortion theory, generalization error
\end{IEEEkeywords}

\section{Introduction}
Learning under communication constraints has received increased attention recently, for instance for distributed learning and sensor networks applications~\cite{strinati20216g}.
When considering a rate-limited channel, one key question is whether the design principles for the encoder and decoder for a learning task still align with those of traditional communication systems, where the main goal is data reconstruction.

To address this issue, researchers have explored simple distributed learning problems involving two correlated sources $X$ and $Y$, where $X$ is the source to be encoded and $Y$ serves as side information at the decoder.
Distributed hypothesis testing has been extensively studied for specific hypothesis tests on the joint distribution $P_{XY}$, and asymptotic limits on the Type-II error exponent have been determined in~\cite{katz2017,salehkalaibar2019,sreekumar2020}. 
Furthermore,~\cite{el2015slepian} demonstrated that the rate required for estimating a parameter $\theta$ from the joint distribution $P_{XY}$ is less than the rate necessary for reconstructing the source. 
 Finally,~\cite{raginsky} developed a universal achievable bound on the learning generalization error, applicable to a wide range of distributed learning problems involving two sources. 
 However, it was later shown in~\cite{jwei23} that this bound is quite loose when applied to linear regression.
Building upon~\cite{jwei23}, this paper focuses on the wider problem of polynomial regression and aims to establish achievable generalization error bounds that improve over the ones presented in~\cite{raginsky}.
Despite its simplicity, polynomial regression,  captures essential learning theory concepts 
and is widely applied in signal and image processing, \emph{e.g.},~\cite{Siggiridou2021,Finlayson2015}.


Morever, this paper investigates a secondary, yet significant concern, which is the trade-off between data reconstruction and learning under communication constraints. 
In this matter,~\cite{blau2019rethinking} demonstrated that there indeed exists a tradeoff between data reconstruction and visual perception. 
Similar tradeoffs have been observed for other problems, such as hypothesis testing in~\cite{katz2017}, or identifying noisy data in a database in~\cite{Tuncel2014}.
All previous works utilize distortion as the figure of merit for data reconstruction and employ distinct measures for the learning aspect; like a divergence between two distributions in~\cite{blau2019rethinking}, and the type-II error exponent in~\cite{katz2017}. Unfortunately, none of these metrics are applicable to polynomial regression, underlining the need for a different analysis in our case. 

Least squares regression, a fundamental statistical prediction problem, has been extensively investigated in literature. 
The ordinary least squares (OLS) estimator is a popular regression method, and its generalization error with $k$ predictors and $n$ samples is known to scale 
as $\frac{k}{n-k+1}$~\cite{Mourtada_2022}.
However, this result does not take into account the communication constraint, which is an important consideration in many practical scenarios. 
%
%
%
In the context of polynomial regression, this paper determines the minimum achievable source coding rate under a constraint on the generalization error for both asymptotic and non-asymptotic regimes. The regions are derived using both standard asymptotic information theory tools~\cite{wyner1976rate,draper2004universal} and finite-length tools~\cite{watanabe2015nonasymptotic}, and they improve over the bounds established by~\cite{raginsky}. Additionally, the analysis reveals that no trade-off exists between data reconstruction and polynomial regression in terms of coding rate.

The outline of the paper is as follows. Section~\ref{problem_statement} defines the problem of coding for polynomial regression. Section~\ref{asymptotic_bound} introduces the asymptotic rate-loss bounds. Section~\ref{non_asymptotic_bound} provides the rate-loss bounds in finite blocklength. Section~\ref{numerical_results} shows numerical results.

\section{Problem statement}
\label{problem_statement}

\subsection{Notation}
Throughout this article, random variables and their realizations are denoted with capital and lower-case letters, respectively, \emph{e.g.}, $X$ and $x$. Random vectors of length $n$ are denoted $\vec{X} = \left[X_1, ..., X_n\right]^T$, and  
$\mathbb{E}[\vec{X}]$ and $\Cov{\vec{X}}{}$ are the expected value and the covariance matrix of $\vec{X}$, respectively. 
Next, $\mat{X} = \left[\vec{X}_1, \cdots, \vec{X}_p \right]$ is a matrix gathering a $p$-length sequence of random vectors $\vec{X}_i$, $i\in \llbracket 1,p\rrbracket$.
We use Tr($\mat X$) to denote the trace of matrix $\mat X$, while  $\lambda_{\max}(\mat X)$ and $\lambda_{\min}(\mat X)$  are  the maximum and minimum eigenvalues of matrix $\mat{X}$,  respectively. We further denote $||\mat{X}||$ as the norm-2 of a matrix $\mat{X}$.
Sets are denoted with calligraphic fonts,  and if $f: \set{X} \rightarrow \set{Y}$ is a mapping then $\left|f\right|$ denotes the cardinality of $\set{Y}$.  Finally $\log(\cdot)$ denotes the base-2 logarithm.

\subsection{Source definitions}
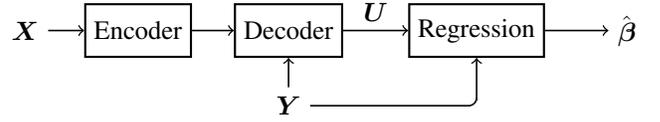
\begin{figure}
   \centering

\begin{tikzpicture}
    \node (X) at (0,-2) {$\vec{X}$};
    \node (beta) at (8,-2) {$\hat{\vec{\beta}}$};
    \node[squarenode,line width=0.8pt] (encoder) at (1.5,-2) { Encoder};
    \node[squarenode,line width=0.8pt] (decoder) at (3.5,-2) { Decoder};
    \node[squarenode,line width=0.8pt] (est) at (6,-2) { Regression};
    \node (Y) at (3.5, -3) {$\vec{Y}$};

    
    \draw[->,line width=0.6pt] (X) -- (encoder.west);
    \draw[->,line width=0.6pt] (encoder) -- (decoder.west);
    \draw[->,line width=0.6pt] (decoder) -- node[above] {$\vec{U}$} (est.west);
    \draw[->,line width=0.6pt] (est) -- (beta);
    \draw[->,line width=0.6pt] (Y) -- (decoder.south);
    \draw[->,rounded corners=3pt,line width=0.6pt] (Y.east) -| (6,-3) -- (est.south);
\end{tikzpicture}
    
    \caption{Coding scheme for regression}
    \label{fig:test_channel_wz}
\end{figure}

Let $(X,Y)\sim P_{XY}$ be a pair of jointly distributed random variables, where $X$ is the source to be encoded and $Y$ is the side information only available at the decoder, see Figure~\ref{fig:test_channel_wz}. For simplicity and without loss of generality, we consider $\Exp{Y}{} = 0$.
We define $\vec{\beta} = [\beta_0, \beta_1, ..., \beta_{k-1}]^T \in \mathbb{R}^k$, and $\vec{Y}^{\star} = [Y^0, Y^1, ..., Y^{k-1}]^T \in \mathbb{R}^k$, where $Y^i$ is the variable $Y$ raised to power $i$. We assume that $X$ follows a polynomial model of order $k$ defined as
\begin{equation}
\label{pl_regression}
X  = \sum_{i=0}^{k-1} \beta_i Y^i + N = \vec{\beta}^T \vec{Y}^{\star} + N ,
\end{equation}
where $N\sim \mathcal{N}(0, \sigma^2)$ follows a Gaussian distribution with mean $0$ and variance $\sigma^2$. The vector $\vec{\beta}$ is constant and unknown at the transmitter. 

\subsection{Polynomial Regression}\label{sec:poly_reg}
Polynomial regression aims at estimating the parameter vector $\vec{\hat{\beta}}$ from realizations, or noisy realizations, of $X$ and $Y$. As a standard supervised learning problem, polynomial regression consists of two phases. We use $X$, $Y$ to denote symbols generated at the training phase, and $\tilde{X}$, $\tilde{Y}$ for symbols generated at the inference phase. The training phase consists of estimating $\vec{\beta}$ from a training sequence composed by the available side information $\vec{Y}$ and by a coded version of $\vec{X}$ which is denoted $\vec{U}$.  The inference phase consists of calculating estimates of the symbols $\tilde{X}$ as $\hat{X} = \hat{\vec{\beta}} \vec{\tilde{Y}}^{\star}$, where $\vec{\hat{\beta}}$ is the estimate of $\vec{\beta}$ from the training phase. Note that the inference phase does not need any data transmission, since the side information $\tilde{Y}$ is directly available to the decoder. 

Following the notation introduced by Raginsky in~\cite{raginsky}, we next formalize the problem as follows. 
Let $\mathcal{F}$ be the set of polynomial functions $f : \mathbb{R} \rightarrow \mathbb{R}$ of the form $f(y) = \vec{\alpha}^T\vec{y}^{\star}$, where $\vec{\alpha} \in \mathbb{R}^k$. Polynomial regression outputs a sequence of functions $\widehat{f}^{(n)} \in \mathcal{F}$, called predictors, such that $\widehat{f}^{(n)} : \set{Z}^n \times \mathbb{R} \rightarrow \mathbb{R} $, where $\vec{Z} = (\vec{U},\vec{Y}) \in \set{Z}^n$ is a training sequence in which $\vec{U}$ and $\vec{Y}$ are  sequences of length $n$. 
Given that $\widehat{f}^{(n)} \in \mathcal{F}$, we can equivalently write
\begin{equation}
\widehat{f}^{(n)}(\vec{Z},y) = \vec{\alpha} (\vec{Z})^T \vec{y}^{\star},
\end{equation} 
where $\vec{\alpha}: \mathcal{Z}^n \rightarrow \mathbb{R}^k$.

Consider the quadratic loss function $\ell: \mathbb{R}^2 \rightarrow \mathbb{R}$ defined as $\ell(x,\hat{x}) = (x - \hat{x})^2 $. 
The minimum expected loss is defined as in~\cite{jwei23, raginsky} as\footnote{One may also define a loss over a sequence. However, since the samples from the training and inference phases are i.i.d. it does not change the analysis.} 
\begin{equation}
\label{minimum_expected_loss}
 L^{\star}(\mathcal{F},\vec{\beta}) = \inf_{f \in \mathcal{F}} \Exp{\ell(X,f(Y))}{}. 
\end{equation}
The generalization error is defined as
\begin{equation}
\label{generalization_error}
     G(\widehat{f}^{(n)},\vec{\beta}) = \Exp{\ell\left( \tilde{X},\widehat{f}^{(n)}(\vec{Z},\tilde{Y}) \right ) | \vec{Z}}{\tilde X \tilde Y} .
 \end{equation}
where $(\tilde{X},\tilde{Y})\sim P_{XY}$ is independent from $\vec{Z}$, the training sequence. The generalization error being a random variable due to the conditioning on $\vec{Z}$, the quantity $\Exp{G\left( \widehat f^{(n)} , \vec{\beta}\right)}{\vec{Z}}$ is referred to as the expected generalization error.

In the previous expressions, the minimum expected loss~\eqref{minimum_expected_loss} simply expresses the average gap between $X$ and $f(Y)$, for the function $f$ that minimizes the quantity $\Exp{\ell(X,f(Y))}{}$ over the space of polynomial functions $\mathcal{F}$. However, there is no guarantee that this optimal function $f$ can be obtained from training. On the opposite, the generalization error measures the learning performance as the expected loss for a certain training sequence $\vec{Z}$. This training sequence allows to produce an estimated function $\widehat{f}^{(n)}(\vec{Z}, \cdot)$ which can then be used to evaluate new samples $\hat{\tilde{X}} = \widehat{f}^{(n)}(\vec{Z},\tilde{Y})$ at the inference phase. Especially, it is easy to show that $\Exp{G\left( \widehat f^{(n)} , \vec{\beta}\right)}{\vec{Z}} \geq  L^{\star}(\mathcal{F},\vec{\beta})$.
Therefore, the gap $ \Exp{G\left( \widehat f^{(n)} , \vec{\beta}\right)}{\vec{Z}} - L^{\star}(\mathcal{F},\vec{\beta})  $ is a key quantity to characterize the performance of a coding scheme dedicated to learning, and this is why our rate-learning regions will be expressed from this quantity.

\subsection{Coding scheme}
The coding scheme is analogue to the one for linear regression in \cite{jwei23}. However, the theoretical analysis differs and becomes more complex, as will be described in the next sections. 

\begin{definition}
A polynomial regression scheme at rate $R$ is defined by a sequence $\{(e_n, d_n, R, \hat{f}^{(n)}))\}$ with an encoder  $e_n: \mathcal{X}^n \xrightarrow[]{} \llbracket 1,M_n\rrbracket$
a decoder $d_n : \mathcal{Y}^n \times \llbracket 1,M_n\rrbracket \rightarrow \mathcal{U}^n$
and the learner $t_n: \set{Y}^n \times \set{U}^n \rightarrow \mathcal{F}$ 
such that
\begin{equation}
    \notag
    \mathop{\lim\sup}\limits_{n\rightarrow\infty} \frac{\log M_n}{n} \leq R .
\end{equation}    
\end{definition}



\begin{definition}\label{def:nmleps}
    An $(n, M, l, \varepsilon)$ code for the sequence $\{(e_n, d_n, R, \hat{f}^{(n)})\}$ and $\varepsilon \in (0,1)$ is a code with $|e_n| = M$ such that 
    \begin{equation}
    \prob{G(\widehat{f}^{(n)},\vec{\beta}) \geq l}{ }{} \leq \varepsilon\  \text{and} \ \frac{\log M}{n} \leq R .
    \end{equation}
\end{definition}


\begin{definition}
    For fixed $l$ and blocklength $n$, the finite blocklength rate-loss functions with excess loss $\varepsilon$
    is defined by:
    \begin{equation}
    \begin{aligned}
        R(n, l, \varepsilon) &= \inf_{R} \{\exists \ \  (n, M, l, \varepsilon) \ code \}
    \end{aligned}
    \end{equation}
\end{definition}

\begin{definition} 
     A pair $(R,\delta)$ is said to be achievable 
     if there exists a sequence $\{(e_n, d_n, R, \hat{f}^{(n)})\}$ such that 
\begin{equation}
    \limsup_{n\rightarrow\infty} \Exp{G(\hat{f}^{(n)}, \vec{\beta})}{\vec{Z}} \leq  L^*(\mathcal{F}, \vec{\beta}) +  \delta
\end{equation}
\end{definition}
As discussed in Section~\ref{sec:poly_reg}, the achievable region is defined in terms of gap between $\Exp{G(\hat{f}^{(n)}, \vec{\beta})}{\vec{Z}}$ and $L^*(\mathcal{F}, \vec{\beta})$. 
Although the regions defined in this section pertain to rate-generalization error regions, for the sake of simplicity and with a minor deviation in terminology, we refer to them as rate-loss regions in the subsequent discussions.

\section{Asymptotic Bound on the Rate-Loss function}
\label{asymptotic_bound}

In~\cite[Theorem 3.3]{raginsky}, it is shown that, for a quadratic loss function, the generalization error can be bounded as:
\begin{equation}
\label{eq:bounds_gen}
\begin{aligned}
       L^{\star \frac{1}{2}}(\mathcal{F},\vec{\beta}) \leq  \limsup_{n\rightarrow\infty}\Exp{G(\hat{f}^{(n)}, \vec{\beta})^{\frac{1}{2}}}{} &\leq L^{\star \frac{1}{2}}(\mathcal{F},\vec{\beta})\\
       &+ 2\mathbb{D}_{X|Y}(R)^{1/2}
\end{aligned}
\end{equation}
where $\mathbb{D}_{X|Y}(R)$ is the conditional distortion-rate function. 
It can be shown that for the polynomial regression, the minimum expected loss in~\eqref{minimum_expected_loss} is $L^{\star}(\set{F},\vec{\beta}) = \sigma^2$. 
In this section, we build a coding scheme that allows to improve the upper bound in~\eqref{eq:bounds_gen} for the polynomial model.  

\subsection{Rate-loss region}

\begin{theorem}
\label{achievability_asymptotic}
    Given any rate $R > 0$, the pair $(R, 0)$ is achievable for the polynomial regression scheme with squared loss, for sources $(X,Y)$ following the polynomial model~\eqref{pl_regression}.
\end{theorem}
This result states that the minimum generalization error which is given by the loss function $L^{\star}(\mathcal{F},\vec{\beta})$ in~\eqref{eq:bounds_gen} can be achieved with any arbitrary rate $R$, as long as the training sequence is long enough.  The proof of the Theorem is based on an achievability scheme built on a Gaussian test channel. This test channel is known for being optimal for joint Gaussian sources when considering data reconstruction~\cite{wyner1978}, although it may be suboptimal for other models like the one we consider in this paper. However, in our case, we show that this test channel achieves the optimal rate-loss region $(R,0)$ for polynomial regression, and we further discuss its optimality for data reconstruction in Section \ref{sec:tradeoff}.

\subsection{Proof of Theorem \ref{achievability_asymptotic} : Achievability scheme}

Let us consider the test channel $U = \alpha(X + \Phi)$, where $\Phi \sim \mathcal{N}(0, \sigma^2_{\Phi})$ is independent of $X$, and $\alpha$ and $\sigma_\Phi^2$ are two parameters which depend on the distribution of $X$ and $Y$. 

The parameters $\vec{\beta}$ and the joint distribution $P_{XY}$ are unknown to the encoder and decoder but the noise variance of the model, i.e. $\sigma^2$, is assumed to be known at the encoder. Hence, the transmission rate is perfectly known at the encoder and the variable-rate scheme in \cite{draper2004universal} becomes a fixed-rate coding scheme in our setup. The same idea of binning is used and the de-binning is performed based on the empirical mutual information between $\vec{x}$ and $\vec{u}$ evaluated thanks to the type of $\vec{x}$ transmitted in a prefix transmission \cite{draper2004universal}. 
Given that $D<\sigma_x^2$ and $(X\left|\right. Y)$ is Gaussian, we show that the rate-distortion function $R_b(D) = \frac{1}{2} \log \left(1 + \frac{\sigma^2}{\sigma_\Phi^2}\right)$ is achievable for $\Exp{d(X, U)}{XU} \leq D$, where $D$ is a function of $\sigma_{\Phi}^2$.



Then, for a training sequences $(\vec{y}, \vec{u})$, the OLS estimator $\hat{\vec{\beta}}$ is given by\cite[Chapter 7]{rencher2008linear}
\begin{align}
    \vec{\hat{\beta}} = \alpha^{-1}(\mat{Y}^\star{\mat{Y}^{\star}}^T)^{-1}\mat{Y}^\star\vec{u}.
\end{align}
where $\mat{Y}^\star = [\vec{Y^\star_1}, ..., \vec{Y^\star_n}] \in \mathbb{R}^{k\times n}$ and this estimator has the following statistical properties : 
\begin{align}
    \Exp{\vec{\hat{\beta}}}{} = \vec{\beta}\ \text{ and } \ \Cov{\vec{\hat{\beta}}|\vec{Y}}{} = \frac{1}{\alpha^2}\sigma_{U|Y}^2 (\mat{Y}^\star{\mat{Y}^\star}^T)^{-1}
\end{align}
where $\Cov{\vec{\hat{\beta}}|\vec{Y}}{}$ is the covariance matrix of $\vec{\hat{\beta}}$ given $\vec{Y}$. 
Hence, the generalization error \eqref{generalization_error} can be rewritten as
\begin{equation}
    \begin{aligned}
        G(\widehat{f}^{(n)},\vec{\beta}) 
        &= \Exp{[\vec{\beta} - \vec{\hat{\beta}} ]^T\vec{\tilde Y^\star} \vec{\tilde Y^\star}^T[\vec{\beta} - \hat{\vec{\beta}}] + \vec{N}^T\vec{N}|\vec Z}{\tilde X\tilde Y} \\
        &=  [\vec{\beta} - \vec{\hat{\beta}} ]^T\Exp{\vec{\tilde Y^\star} \vec{\tilde Y^\star}^T}{\tilde Y}[\vec{\beta} - \vec{\hat{\beta}} ] + \sigma^2 .
    \end{aligned}
\end{equation}
Let $\mat{\tilde{\Sigma}} = \Exp{\vec{\tilde Y^\star} \vec{\tilde Y^\star}^T}{\tilde Y}$ and $\mat{\Sigma} = \frac{1}{n}\mat{Y}^\star{\mat{Y}^\star}^T$.
Then, the expected generalization error is
\begin{align}
\notag
    &\Exp{G(\widehat{f}^{(n)},\vec{\beta})}{\vec Z} \\ \notag
    &= \sigma^2 + \Exp{\frac{1}{n}(\mat{\Sigma}^{-1}\mat{Y}^\star(\vec N+\vec\Phi))^T \mat{\tilde{\Sigma}}\frac{1}{n}(\mat{\Sigma}^{-1}\mat{Y}^\star(\vec N+\vec\Phi))}{}\\
    &=\sigma^2 + \frac{\sigma^2 + \sigma_\Phi^2}{n}\Exp{\text{Tr}\left(\mat{\Tilde{\Sigma}}\mat{\Sigma}^{-1}\right)}{}\label{expectation_g_e} .
\end{align}
The next step is to show that $\Exp{\text{Tr}\left(\Tilde{\mat \Sigma}\mat \Sigma^{-1}\right)}{}$ is bounded by some constant $C$ for $n$ large enough. The following proposition bounds the trace of a product of two matrices by their eigenvalues.
\begin{proposition}{\cite[p 340]{marshall11}}
\label{ry's inequality}
(Ruhe’s trace inequality). If $\mat{U}$ and $\mat{V}$ are $k\times k$ positive semidefinite Hermitian matrices with eigenvalues $\lambda_i(\mat{U}), \lambda_i(\mat{V})$, $i \in \left\{1,\cdots, k\right\}$ then
\begin{align}
   \mathrm{Tr}(\mat{U} \mat{V})\leq \sum_{i=1}^k \lambda_i(\mat{U})\lambda_i(\mat{V})
\end{align}
\end{proposition}
\begin{lemma}
\label{lemma_min_eigenvalue}
    If $\mat{A}$ and $\mat B$ are real symmetric matrices, then:
    \begin{align}
        \lambda_{\min}(\mat{A}) \geq \lambda_{\min}(\mat{B}) - ||\mat{A} - \mat{B}||
    \end{align}
\end{lemma}
\begin{IEEEproof}
    Let $\vec{x}$ be a vector such that $\left|\left| x \right|\right|_2=1$, by Cauchy-Schwartz inequality, for a real symmetric matrix $\mat{M}$, we have 
    \begin{align}
        -||\mat M||\le \vec{x}^T\mat{M}\vec{x} \le ||\mat{M}||.
    \end{align}
    With the properties of eigenvalues, we have 
    \begin{align}
        \lambda_{\min}(\mat{M}) \leq \vec{x}^T\mat{M}\vec{x} \leq \lambda_{\max}(\mat{M}) .
    \end{align}
    For real symmetric matrices $\mat{A}$ and $\mat B$, we have
    \begin{align}
        \vec{x}^T\mat{A}\vec{x} = \vec{x}^T\mat{B}\vec{x} + \vec{x}^T(\mat{A}-\mat{B})\vec{x} .
    \end{align}
    Applying the above inequalities shows the desired result.
\end{IEEEproof}
We remark that $\mat\Sigma$ is an estimator of the covariance matrix of $\vec{Y}$. Then, from Proposition \ref{ry's inequality} and Lemma \ref{lemma_min_eigenvalue}, for $n$ large enough, $\text{Tr}\left(\Tilde{\mat \Sigma}\mat \Sigma^{-1}\right)$ is bounded almost surely by:
\begin{align}
        \text{Tr}\left(\Tilde{\mat \Sigma}\mat \Sigma^{-1}\right) 
        &\leq k \frac{\lambda_{\max}(\mat{\tilde{\Sigma}})}{\lambda_{\min}(\mat{\tilde \Sigma}) - ||\mat{\tilde \Sigma} - \mat{ \Sigma}||} .
\end{align}
Substituting this into \eqref{expectation_g_e} with some constant $C = \frac{\lambda_{\max}(\mat{\tilde{\Sigma}})}{\lambda_{\min}(\mat{\tilde \Sigma})}$ and the fact that $||\mat{\tilde \Sigma} - \mat{ \Sigma}||\rightarrow 0$ almost surely, shows that the expected generalization error is upper bounded by
\begin{equation}
    \label{eq_generalization_error}    \Exp{G(\widehat{f}^{(n)},\vec{\beta})}{\vec Z} \leq \sigma^2 + \frac{(\sigma^2 + \sigma_\Phi^2)}{n} kC
\end{equation}
Thus $\Exp{G(\widehat{f}^{(n)},\vec{\beta})}{\vec Z} \rightarrow \sigma^2$ as $n \rightarrow \infty$, which completes the proof. 

Our result closes the gap between the lower bound and the upper bound from~\cite{raginsky} (see equation~\eqref{eq:bounds_gen}). In order to provide a bound applicable to a wide range of problems, the upper bound from~\cite{raginsky} considered both the observation noise between $\vec{X}$ and $\vec{Y}$ and the distortion between $\vec{X}$ and $\vec{U}$. While in our result, by the Gaussian test channel and OLS estimation from $\vec{U}$ and $\vec{Y}$, we show that the quantification error term in \eqref{eq_generalization_error}, and hence the distortion term, is vanishing with the block-length~$n$.

\subsection{Trade-off between data reconstruction and polynomial regression}\label{sec:tradeoff}
In this section, we show that the previous achievability scheme considered for polynomial regression also achieves the optimal Wyner-Ziv rate-distortion function for data reconstruction, for sources modeled by \eqref{pl_regression}.
\begin{corollary}
    For a pair of sources $(X,Y)$ modeled from~\eqref{pl_regression}, there is no trade-off in terms of coding rate between distortion and polynomial regression generalization error.
\end{corollary}
\begin{IEEEproof}
We first investigate the conditional setup in which the side information $Y$ is also available at the encoder. Since the random variable $(X|Y) \sim \mathcal{N}(0,\sigma^2)$, the following conditional rate-distortion function can be achieved~\cite{gray1972conditional}
\begin{equation}\label{eq:rd_cond}
    R_{X|Y}(D) = \frac{1}{2}\log \left( \frac{\sigma^2}{D} \right) ,
\end{equation}
where $D = \Exp{(X-\hat{X})^2}{}$ is the distortion. 
We now show that in the Wyner-Ziv setup where $Y$ is only available at the decoder, the rate-distortion function $R_{\text{WZ}}(D)$ is equal to $R_{X|Y}(D)$ when considering the same test channel $U = \alpha (X+\Phi)$ as in the proof of Theorem~\ref{achievability_asymptotic}, with $\alpha = \frac{\sigma^2 - D}{\sigma^2}$, and $\sigma^2_{\Phi} = \frac{D\sigma^2}{\sigma^2 - D}$. By using the proposed achievability scheme, the random variable $U$ can be recovered perfectly at the decoder, and then produces $\hat{X} = U + (1-\alpha)\vec{\beta}^T \vec{Y}^\star$. This allows us to evaluate $\Exp{(X-\hat{X})^2}{}  = (\alpha-1)^2\sigma^2 + \alpha^2 \sigma_{\Phi}^2$.
Replacing $\alpha$ and $\sigma_{\Phi}^2$ by their expressions leads to $\Exp{(X-\hat{X})^2}{} = D$.
Second, the Wyner-Ziv rate-distortion function has expression~\cite{wyner1978}
\begin{align}\notag
    I(X;U) - I(Y;U) 
    & = \frac{1}{2} \log_2\left(\frac{\sigma^2 + \sigma_{\Phi}^2}{\sigma_{\Phi}^2}\right)
\end{align}
where the equality comes from the fact that $N$ and $\Phi$ are Gaussian random variables. 
Replacing $\sigma_{\Phi}^2$ by its expression gives that $R_{\text{WZ}}(D) = R_{X|Y}(D)$ in~\eqref{eq:rd_cond}, which shows that the Gaussian test channel is optimal when considering our polynomial source model. 
Note that in the previous derivation, we considered that $\boldsymbol{\beta}$ is perfectly known. If this is not the case, $\hat{X}$ is computed from $\hat{\boldsymbol{\beta}}$ instead of $\boldsymbol{\beta}$, and following the same derivation as for the generalization error permits to show that $\Exp{(X - \hat{X})^2}{} \rightarrow D$ as $n\rightarrow \infty$. 

This result differs from the other ones in literature that show that there is a tradeoff between reconstruction and learning, such as for the hypothesis testing problem for instance~\cite{katz2017}.




\end{IEEEproof}



\section{Rate-Loss non-asymptotic bound}
\label{non_asymptotic_bound}

In the finite-blocklength regime, not all codewords satisfy the generalization error constraint, and hence the excess probability, defined in Definition \ref{def:nmleps}, has to be taken into account. The characterization of the non-asymptotic achievable bound for the rate-generalization error region is built from the rate-distortion problem in finite blocklength regime, studied in \cite{watanabe2015nonasymptotic}. Similarly, we define the information-loss density vector as follows:
\begin{align} \label{eq:informationvector}
    \vec {i}(U,X,Y, \Tilde{X}, \tilde{Y}):=\begin{bmatrix}\textstyle -\log \displaystyle \frac {P_{U|Y}(U|Y)}{P_{U}(U)}\\[9pt]\textstyle \log \displaystyle \frac {P_{U|X}(U|X)}{P_{U}(U)}\\[9pt]\textstyle \ell(\Tilde{X}, \hat{f}^{(n)}(\vec{Z},\Tilde{Y})) \end{bmatrix}\!
\end{align}
where 
the third term is specific to our non-linear regression problem. 
The expectation of $\vec{i}$ over the distribution $P_{UXY\tilde{X}\tilde{Y}}$ is $\vec {J} = \left[-I(U;Y), I(U;X), \Exp{G(\hat{f}^{n}, \vec{\beta})}{\vec Z}\right]^T$,
where the sum of the first two components gives the Wyner-Ziv coding rate. The covariance matrix of \eqref{eq:informationvector} is
\begin{align}
    \vec{V} = \Cov{\vec{i}(U,X,Y, \tilde{X}, \tilde Y)}{} .
\end{align}
Let $k$ be a positive integer and $\vec{V}\in\mathcal{R}^{k\times k}$ be a positive-semi-definite matrix. Given a Gaussian random vector $\vec{B}\sim \mathcal{N}(0,\vec{V})$, the dispersion region is \cite{tan_three_nit}
\begin{align}
\label{dispersion}
    \mathscr {S}( \vec {V}, \varepsilon ):= \{ \vec {b}\in \mathbb {R}^{k}: \Pr ( \vec {B}\le \vec {b})\ge 1- \varepsilon \}.
\end{align}

By replacing the distortion measure by the generalization error, and adapting some steps of the analysis, we can obtain a similar result as Theorem 2 in \cite{watanabe2015nonasymptotic}. Finally, by applying this theorem in conjunction with the multidimensional Berry-Esséen Theorem, we show that for all $0<\varepsilon<1$ and $n$ sufficiently large, the $(n , \varepsilon)$-rate-generalization error function satisfies:
\begin{align}\label{non-asym-r}
        R_{ \mathrm {b}}(n, \varepsilon,l)\le&\inf \bigg \{\vec{M}^T \left ({ \vec {J}+ \frac { \mathscr {S}( \vec {V}, \varepsilon ) }{\sqrt {n}} + \frac {2\log n}{n} \vec {1}_{3}}\right )\bigg \}
\end{align}
with $\vec{M} = [1\quad 1\quad 0]^T$ and $\vec{1}_3 = \left[1~~1~~1\right]^T$.

\section{Numerical results}
\label{numerical_results}
\begin{figure}
    \centering
    \includegraphics[width=0.9\columnwidth]{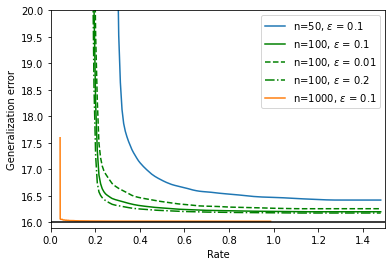}
    \caption{Non-asymptotic rate-generalization error region labeled on the blocklength $n$ and the excess loss probability $\varepsilon$.}
    \label{fig:rate-error}
\end{figure}

Let us consider $X = \beta_0 + \beta_1 Y + \beta_2 Y^2 + N$, and assume that $Y$ is uniform over $\left[-1 , 1\right]$. We also set $\vec{\beta} = [2, 3, 1]^T$ and $\sigma^2 = 16$. From the theorem of change variable, for $\beta_2 > 0$ and $\beta_1^2 + 4\beta_2(v-\beta_0)\geq0$, the distribution of $V=\vec{\beta}^T \vec{Y}^\star$ is: 
\begin{equation}
    P_V(v) = \begin{cases}
        \frac{1}{\sqrt{\beta_1^2 + 4\beta_2(v-\beta_0)}} & |y_1(v)| \le 1 \ \text{and}\  |y_2(v)| \le 1\\
        \frac{1}{2\sqrt{\beta_1^2 + 4\beta_2(v-\beta_0)}}  & |y_1(v)| \le 1 \ \text{or}\  |y_2(v)| \le 1\\
        0 &  \text{otherwise}
    \end{cases}\notag
\end{equation}
where $y_1 = \frac{-\beta_1 - \sqrt{\beta_1^2 + 4\beta_2(v-\beta_0)}}{2\beta_2}, y_2 = \frac{-\beta_1 + \sqrt{\beta_1^2 + 4\beta_2(v-\beta_0)}}{2\beta_2}$. The probability density function of $U = \alpha (V + N + \Phi)$ can then be expressed as 
\begin{align}
    P_U(u) = \frac{1}{\alpha\sqrt{2\pi(\sigma^2 + \sigma_\Phi^2)}}\int_{-\infty}^{\infty} P_V(v) e^{-\frac{(\frac{u}{\alpha}-v)^2}{2(\sigma^2+\sigma_\Phi^2)}}dv
\end{align}
which can be evaluated numerically. Using \eqref{non-asym-r} with $(U|Y) \sim \mathcal{N}(0, \alpha^2(\sigma^2 + \sigma_\Phi^2))$ and $(U|X) \sim \mathcal{N}(0, \alpha^2\sigma_\Phi^2)$, we can estimate the information-density-loss vector by generating a large number of samples, and thus estimate the dispersion region in \eqref{dispersion}. 
Figure \ref{fig:rate-error} shows the boundaries of the achievable rate-loss region for different parameters $n$ and $\varepsilon$. The black line represents the best achievable generalization error, i.e. $\sigma^2$. We observe that the achievable region enlarges when the source size, $n$, or the excess probability increases. Indeed, when the excess probability is larger, the proportion of codewords which exceeds the generalization error constraint is larger, and this situation occurs for smaller rate. Moreover, for a fixed excess probability, increasing $n$ allows to reduce the rate since the poorly reconstructed $U$ is compensated by the large number of samples for estimating the regression parameters. These results do not deal with an outer bound at finite blocklength, i.e. a rate-loss region that \emph{cannot} be exceeded, and the region outside the boundary needs further investigation. 


\section{Conclusion} \label{conclusion}
This paper provided achievable rate-generalization error regions for the polynomial regression problem in both asymptotic and non-asymptotic regimes. 
An important result of our study states that asymptotically there is no trade-off between data reconstruction and polynomial regression under communication constraints. 
The characterization of the outer bound (converse) for the rate-generalization error region is also of great interest and would allow to refine the analysis. The developed framework could be extended to more complex learning taks, such as non-parametric estimation, in the future.



\bibliographystyle{IEEEtran}
\bibliography{draft}

\end{document}